# A Cartesian Quasi-classical Model to Nonequilibrium Quantum Transport: The Anderson Impurity Model

Bin Li,[a] Tal J. Levy,[b] David W.H. Swenson,[c] Eran Rabani,[b] and William H. Miller[a]

*(a) Department of Chemistry and Kenneth S. Pitzer Center for Theoretical Chemistry, University of California, and Chemical Sciences Division, Lawrence Berkeley National Laboratory, Berkeley, California 94720, USA*
*(b) School of Chemistry, The Sackler Faculty of Exact Sciences, Tel Aviv University, Tel Aviv 69978, Israel*
*(c) van 't Hoff Institute for Molecular Sciences, Universiteit van Amsterdam, PO Box 94157, 1090 GD Amsterdam, The Netherlands*

Abstract: We apply the recently proposed quasi-classical approach for a second quantized many-electron Hamiltonian in Cartesian coordinates [*J. Chem. Phys.* **137**, 154107 (2012)] to correlated nonequilibrium quantum transport. The approach provides accurate results for the resonant level model for a wide range of temperatures, bias and gate voltages, correcting for the flaws of our recently proposed mapping using action-angle variables. When electron-electron interactions are included, higher order schemes are required to map the two-electron integrals, leading to semi-quantitative results for the Anderson impurity model. In particular, we show that the current mapping is capable of capturing qualitatively the Coulomb blockade effect.

## I. INTRODUCTION

The growing interest in the properties of molecular electronic devices has raised fundamental and conceptual issues regarding the physics of nanometer scale systems.[1-3] Theory faces several important challenges: (a) understanding the coupling of individual molecules to macroscopic electrodes under nonequilibrium conditions,[4,5] (b) the characterization of the temperature dependence of conductance as well as the role of molecular vibrations, environmental disorder and dissipation are of crucial importance,[6] (c) electron-electron correlations, that are usually neglected in macroscopic electrodes, by the application of an effective band-structure description of the leads, are very important within the molecule or quantum dot and thus strongly affect the conduction.[7,8] Finally, (d) characterizing transport behavior in systems driven by weak and strong electromagnetic fields and the optical properties of such junctions is another difficult issue which must be overcome by theory.[9]

Generally, transport through nanoscale interacting systems is a many-body out of equilibrium problem that cannot be solved exactly, but for a few simple cases.[10-13] Several theoretical methods have been proposed to address the transport for such systems, and the different approaches can be loosely classified as numerically exact,[14-21] perturbative,[9,22,23] and phenomenological.[24-32] While successful, these approaches suffer from several limitations. In particular, the more advanced techniques cannot be extended to cases of more complex environments, an extension required to adequately describe transport through molecules.

Parallel to these developments, another paradigm has been formulated based on a semiclassical description of the dynamics of a many-body quantum system.[33-35] Recently, Swenson et al.[13,36] have adopted a semiclassical mapping approach based on the early work of Miller and White,[37] constructing a classical Hamiltonian corresponding to the general second-quantized Hamiltonian operator for a many-electron system in which all the creation and annihilation operators for the spin-orbitals were substituted by a set of classical action-angle variables. This scheme was employed to calculate the current of the resonant level model[13] and for an inelastic tunneling Holstein model,[36] ignoring in both cases electron-electron interactions. The mapping approach provides semi-quantitative results for a range of system parameters, i.e. different source-drain voltages, temperatures and electron-phonon couplings. However, it falls short in providing accurate and even qualitative results in several different limits. For example, it fails to capture the effects of a gate voltage in the noninteracting case,[13] or the well-known inelastic tunneling peaks when the phonon bath is characterized by a sharp spectral function.[36] Moreover, it does not fully capture the Coulomb blockade[38] when electron-electron interactions are introduced (see further discussion below).

In this paper we present a different mapping procedure for the many-electron second quantized Hamiltonian which is isomorphic to quaternions, based on the approach described in Ref. 41. The occupation numbers and single-electron coupling operators are mapped to the cross product of coordinate and their conjugate momentum vectors, both represented in Cartesian coordinates. This kind of mapping can naturally keep all the anti-commutative relationship between the fermionic creation and annihilation operators in the classical Hamiltonian. With the Cartesian expression for the Hamiltonian, the semiclassical initial value representation method[34] can also be used to calculate the real time dynamics. For nonequilibrium dynamics, e.g. the noninteracting resonant level model,[39] the approach provides accurate results even when a finite gate voltage is applied, correcting for the flaws of our previous action-angle mapping.[13] Moreover, it qualitatively captures the Coulomb blockade observed for large electron-electron interactions as applied in the Anderson impurity model.[40]

The paper is arranged as follows. In section II we describe the Cartesian model for the second quantized many-electron Hamiltonian, the quasi-classical approximation for the occupation numbers, and provide an expression for the current. Results for the classical calculation for the resonant level model and the Anderson impurity model for the different system parameters (i.e. the source-drain voltages, gate voltages, temperatures and one site repulsion) are described in section III. Section IV provides a summary and conclusions.

## II. MODEL AND MAPPING PROCEDURE

### A. Mapping to Cartesian Hamiltonian for the many-electron system

We describe the mapping procedure for the Anderson impurity model. The extension to more sophisticated models is straightforward following the principles described below. We note in passing that the limit $U \to 0$ corresponds to the resonant level model. The second quantized many-electron Hamiltonian for the Anderson impurity model is given by:[40]

$$H = \sum_{\sigma=\uparrow\downarrow} \varepsilon_\sigma d_\sigma^\dagger d_\sigma + U d_\uparrow^\dagger d_\uparrow d_\downarrow^\dagger d_\downarrow + \sum_{\substack{k \in L,R \\ \sigma=\uparrow\downarrow}} \varepsilon_k c_{k\sigma}^\dagger c_{k\sigma} + \sum_{\substack{k \in L,R \\ \sigma=\uparrow\downarrow}} t_k c_{k\sigma}^\dagger d_\sigma + \text{h.c.} \quad (1)$$

In the above equation, $d_\sigma^\dagger$ ($d_\sigma$) are the creation (annihilation) operators of an electron with spin $\sigma = \uparrow, \downarrow$ and similarly, $c_{k\sigma}^\dagger$ ($c_{k\sigma}$) are the creation (annihilation) operators of an electron in mode $k$ of the leads. $\varepsilon_\sigma$ is the one-body energy of the impurity, $U$ is the onsite two-body repulsion, $\varepsilon_k$ is the energy associated with mode $k$ of the left (L) or right (R) noninteracting lead, and $t_k$ is the coupling between the quantum dot and mode $k$.

Following the recent work of Li and Miller,[41] the anticommutative relationship between the creation and annihilation operators, isomorphic to quaternions, can be mapped to the cross product of the 2-dimension coordinate ($\mathbf{r}$) and conjugate momentum ($\mathbf{p}$):

$$c_{k\sigma}^\dagger c_{k\sigma} = \frac{1}{2} + \frac{\sqrt{-1}}{2} i_{k\sigma} j_{k\sigma} \mapsto \frac{1}{2} + \mathbf{r}_{k\sigma} \times \mathbf{p}_{k\sigma} = \frac{1}{2} + x_{k\sigma} p_{y,k\sigma} - y_{k\sigma} p_{x,k\sigma} \quad (2)$$

and

$$c_{k\sigma}^\dagger d_\sigma + d_\sigma^\dagger c_{k\sigma} = \frac{\sqrt{-1}}{2} (i_{k\sigma} j_\sigma + i_\sigma j_{k\sigma}) \mapsto$$
$$\mathbf{r}_{k\sigma} \times \mathbf{p}_\sigma + \mathbf{r}_\sigma \times \mathbf{p}_{k\sigma} =$$
$$x_{k\sigma} p_{y,\sigma} - y_{k\sigma} p_{x,\sigma} + x_\sigma p_{y,k\sigma} - y_\sigma p_{x,k\sigma}, \quad (3)$$

in which $i_{k\sigma}$ and $j_{k\sigma}$ are the basic elements of quaternions. In this notation, the Anderson impurity Hamiltonian takes the form:

$$H = \sum_{\sigma=\uparrow\downarrow} \varepsilon_\sigma (x_\sigma p_{y,\sigma} - y_\sigma p_{x,\sigma}) + U(x_\uparrow p_{y,\uparrow} - y_\uparrow p_{x,\uparrow})(x_\downarrow p_{y,\downarrow} - y_\downarrow p_{x,\downarrow}) + \sum_{\substack{k \in L,R \\ \sigma=\uparrow\downarrow}} \varepsilon_k (x_{k\sigma} p_{y,k\sigma} - y_{k\sigma} p_{x,k\sigma}) + \sum_{\substack{k \in L,R \\ \sigma=\uparrow\downarrow}} t_k (x_\sigma p_{y,k\sigma} - y_\sigma p_{x,k\sigma} + x_{k\sigma} p_{y,\sigma} - y_{k\sigma} p_{x,\sigma}).$$
(4)

In the above equation, we have subtracted a $\frac{1}{2}$ from the classical mapping of the occupation number operator $\hat{n}_{k\sigma} = c_{k\sigma}^\dagger c_{k\sigma}$ in Eq. (2) to include the Langer correction.

In what follows, we will mainly be interested in the calculation of the time dependent current. In the above notation, the current from the left/right lead to the dot is given by:

$$I_{L(R)} = -\frac{d}{dt} \sum_{\substack{k \in L(R) \\ \sigma=\uparrow\downarrow}} \langle x_{k\sigma} p_{y,k\sigma} - y_{k\sigma} p_{x,k\sigma} \rangle$$
$$= \sum_{\substack{k \in L(R) \\ \sigma=\uparrow\downarrow}} t_k \langle y_\sigma p_{y,k\sigma} + x_\sigma p_{x,k\sigma} - x_{k\sigma} p_{x,\sigma} - y_{k\sigma} p_{y,\sigma} \rangle \quad (5)$$

and the total current is $I = (I_L - I_R)/2$.

## B. Special treatment for the two-body terms

As will become clear below, the above mapping combined with a quasi-classical description of the dynamics fails to capture the well-known Coulomb blockade effect at low temperatures and high values of $U$. This is a result of the fact that within the quasi-classical approximation the occupation numbers assume a continuum value and thus, when the dot is occupied by a "fraction" of an electron with spin up it does not block completely the path of an electron with spin down.

One way to overcome this faulty is to increase the strength of the interactions between electrons of different spins for fractional occupation by taking advantage of the anti-commutation relation of the creation and annihilation operators of fermions $d_\sigma^\dagger d_\sigma + d_\sigma d_\sigma^\dagger = 1$ and the relation $\left(d_\sigma^\dagger d_\sigma\right)^\lambda = d_\sigma^\dagger d_\sigma$, where $\lambda$ is an arbitrary positive integer. Thus, we propose to replace the term $U d_\uparrow^\dagger d_\uparrow d_\downarrow^\dagger d_\downarrow$ in Eq. (1) with $U \left(d_\uparrow^\dagger d_\uparrow\right)^{\lambda_1} \left(d_\downarrow^\dagger d_\downarrow\right)^{\lambda_2}$. Quantum mechanically, this is an exact identity that has no effect on the transport properties of the system or on its thermodynamic state. For the mapped variables, we rewrite the Hamiltonian as:

$$H = \sum_{\sigma=\uparrow\downarrow} \varepsilon_\sigma \left(x_\sigma p_{y,\sigma} - y_\sigma p_{x,\sigma}\right) + \\ U \left(x_\uparrow p_{y,\uparrow} - y_\uparrow p_{x,\uparrow}\right)^{\lambda_1} \left(x_\downarrow p_{y,\downarrow} - y_\downarrow p_{x,\downarrow}\right)^{\lambda_2} + \\ \sum_{\substack{k \in L,R \\ \sigma=\uparrow\downarrow}} \varepsilon_k \left(x_{k\sigma} p_{y,k\sigma} - y_{k\sigma} p_{x,k\sigma}\right) + \\ \sum_{\substack{k \in L,R \\ \sigma=\uparrow\downarrow}} t_k \left(x_\sigma p_{y,k\sigma} - y_\sigma p_{x,k\sigma} + x_{k\sigma} p_{y,\sigma} - y_{k\sigma} p_{x,\sigma}\right),$$

(6)

where $\lambda_1$ and $\lambda_2$ are positive integer. Here, since the mapping is not exact, the results will depend on the choice of $\lambda_1$ and $\lambda_2$. However, we find that the current converges as we increase $\lambda_1$ and $\lambda_2$ to about 20, which is the choice for most of the results reported here (unless otherwise stated).

The physical significance of the special treatment of the two-body terms is simple: Since the occupations $\left(x_\uparrow p_{y,\uparrow} - y_\uparrow p_{x,\uparrow}\right)$ and $\left(x_\downarrow p_{y,\downarrow} - y_\downarrow p_{x,\downarrow}\right)$ can assume any value between 0 and 1, using higher values of $\lambda_{1,2}$ will result in a sharp change of the Coulomb repulsion as the occupations approach the value of 1.

We find numerically that for $\lambda_{1,2} < 20$ the blockade is not significant enough while for larger values the results are numerically converged (see Figure 4 below).

## C. Quasi-classical approximation

In the quasi-classical procedure, the dynamics of the Cartesian variable follow from Hamilton's equations of motions. For the Anderson impurity model, these are (assuming $t_k$ is real):

$$\dot{x}_\sigma = -\varepsilon_\sigma y_\sigma - \sum_{k \in L,R} t_k y_{k\sigma} - U y_\sigma \left(x_{\bar\sigma} p_{y,\bar\sigma} - y_{\bar\sigma} p_{x,\bar\sigma}\right)$$
$$\dot{y}_\sigma = \varepsilon_\sigma x_\sigma + \sum_{k \in L,R} t_k x_{k\sigma} + U x_\sigma \left(x_{\bar\sigma} p_{y,\bar\sigma} - y_{\bar\sigma} p_{x,\bar\sigma}\right)$$
$$\dot{p}_{x,\sigma} = -p_{y,\sigma} \varepsilon_\sigma - \sum_{k \in L,R} t_k p_{y,k\sigma} - U p_{y,\sigma} \left(x_{\bar\sigma} p_{y,\bar\sigma} - y_{\bar\sigma} p_{x,\bar\sigma}\right)$$
$$\dot{p}_{y,\sigma} = \varepsilon_\sigma p_{x,\sigma} + \sum_{k \in L,R} t_k p_{x,k\sigma} + U p_{x,\sigma} \left(x_{\bar\sigma} p_{y,\bar\sigma} - y_{\bar\sigma} p_{x,\bar\sigma}\right)$$
$$\dot{x}_{k\sigma} = -\varepsilon_k y_{k\sigma} - t_k y_\sigma$$
$$\dot{y}_{k\sigma} = \varepsilon_k x_{k\sigma} + t_k x_\sigma$$
$$\dot{p}_{x,k\sigma} = -\varepsilon_k p_{y,k\sigma} - t_k p_{y,\sigma}$$
$$\dot{p}_{y,k\sigma} = \varepsilon_k p_{x,k\sigma} + t_k p_{x,\sigma}$$

(7)

where $\bar\sigma = \uparrow, \downarrow$ for $\sigma = \downarrow, \uparrow$.

To complete the mapping procedure we need to define the initial conditions for all classical phase-space variables. We follow a similar route discussed in Ref. 13 for the action-angle mapping, which recovers the correct statistical behavior (at $t=0$). Since we are interested in a non-correlated initial state with thermally populated leads and an unpopulated quantum dot, we can populate each degree of freedom independently. We enforce quantum statistics on the initial conditions for each degree of freedom by setting the initial occupation to either 0 or 1 such that its expectation value, averaged over the set of initial conditions, satisfies the Fermi-Dirac distribution. Specifically, we choose a random number $\xi_{k\sigma}$ in the interval $[0\cdots1]$, and then select the occupation of mode $k\sigma$ according to:

$$n_{k\sigma} = \begin{cases} 0 & \xi_{k\sigma} > \left(1+e^{\beta(\epsilon_k - \mu_{L/R})}\right)^{-1} \\ 1 & \xi_{k\sigma} \leq \left(1+e^{\beta(\epsilon_k - \mu_{L/R})}\right)^{-1} \end{cases} \quad (8)$$

where $\mu_{L/R}$ is the chemical potential of the left/right lead. The phase space variables are then sampled according to:[42-45]

$$x_{k\sigma} = r_{k\sigma} \cos\theta_{k\sigma}$$
$$p_{xk} = p_{r,k\sigma} \cos\theta_{k\sigma} - n_{k\sigma} \frac{\sin\theta_{k\sigma}}{r_{k\sigma}} \quad (9)$$

in the $x$ direction, and

$$y_{k\sigma} = r_{k\sigma} \sin\theta_{k\sigma}$$
$$p_{yk} = p_{r,k\sigma} \sin\theta_{k\sigma} + n_{k\sigma} \frac{\cos\theta_{k\sigma}}{r_{k\sigma}} \quad (10)$$

in the $y$ direction.

Since the Cartesian mapping introduces 2 classical phase space variables for each creation and annihilation operators, this leads to freedom in choosing the initial conditions, as long as the population given by $n_{k\sigma} = x_{k\sigma} p_{y,k\sigma} - y_{k\sigma} p_{x,k\sigma}$ satisfies Eq. (8). The relations defined by Eqs. (9) and (10) guarantees that this constraint is not violated regardless of the values of $r_{k\sigma}$ and $p_{r,k\sigma}$. In the applications reported below we take $r_{k\sigma} = 1$ and $p_{r,k\sigma} = 0$. This choice provides rapid convergence of the calculated currents with respect to the number of trajectories. Other choices may be based on a Gaussian sampling of the radial position and momentum, etc.

### D. Spectral density and parameters

We adopt a wide band limit to model the spectral density of the leads with a sharp cutoff at high and low energy values:

$$J_{L/R}(\varepsilon_k) = \frac{\Gamma_{L/R}}{\left(1 + e^{A(\varepsilon_k - \frac{1}{2}B)}\right)\left(1 + e^{-A(\varepsilon_k + \frac{1}{2}B)}\right)}. \quad (11)$$

In the above, $B$ is the width spectral function, $\Gamma_{L/R}$ determines the strength of coupling to the left or right leads, and $A$ relates to the sharpness of the cutoff. For the results described below we use $\Gamma_L = \Gamma_R = \frac{1}{2}$, $\Gamma = \Gamma_L + \Gamma_R$, $A = 5\Gamma$, and $B = 20\Gamma$. In terms of the spectral function, the couplings $t_k$ are given by:

$$t_k = \sqrt{\frac{J(\varepsilon_k)\Delta\varepsilon}{2\pi}}, \quad (12)$$

where $\Delta\varepsilon$ is the discretization interval.

In our study of the Anderson impurity model we limit ourselves (without loss of generality) to the common case referred to as the symmetric version, in which the dot states are given by $\varepsilon_\uparrow = \varepsilon_\downarrow = -\frac{1}{2}U$. We study the transient current as a function of temperature $(\beta^{-1} = k_B T)$, source-drain bias $(eV_{SD} = \mu_R - \mu_L)$ and the value of $U = [0, \cdots, 8\Gamma]$.

### III. RESULTS

We solve Hamilton's equation of motion for the mapped variables with initial conditions sampled from the Fermi distribution and the quasi-classical procedure described above. The occupation of the dot is taken to be 0 initially. We use an adaptive time step Runge-Kutta algorithm for the propagation of trajectories. For each lead, 800 spin orbitals are used (400 modes for the $\uparrow$ spin and 400 modes for the $\downarrow$ spin). For most of the results presented below, convergence is achieved with $10^6$ trajectories.

### A. The resonant level model

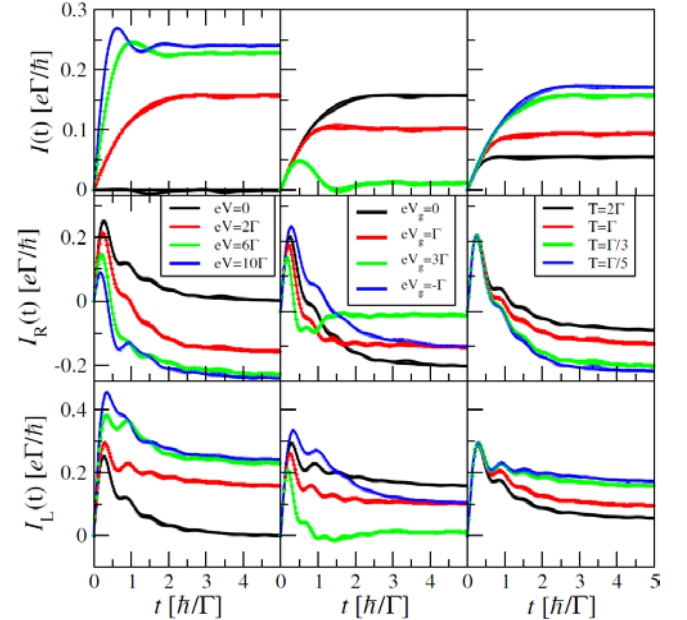

Figure 1: Transient currents for the resonant level model. Left, middle and right panels show results for different source-drain voltages, different gate voltages, and different temperatures, respectively. Lower, middle and upper panels correspond to the left, right and total currents, respectively. Solid lines (hidden by the symbols) are the exact results derived in Ref. 13. Symbols are the results obtained from the Cartesian mapping.

In Figure 1 we plot the left current (lower panels), right current (middle panels), and total current (upper panels) for the case where $U = 0$, corresponding to the resonant level model. Since the different spins do not interact in this limit, we only consider one spin and ignore the other. Left, middle and right panels show the results for different source-drain voltages, gate voltages and temperatures, respectively. The choice of parameters is identical to the case studied by us using the action-angle mapping.[13]

As clearly can be seen, the Cartesian mapping provides accurate results for both the right/left currents and for the total current in comparison to exact results.[13] In fact, the Cartesian mapping is accurate within the statistical error for all range of source-drain and gate voltages studied and for all temperatures studied. The mapping captures the transient currents as well as their steady-state values as time approaches $\Gamma t/\hbar \to 5$.

In comparison to the results obtained using the action-angle mapping presented in Ref. 13, this is a significant improvement, particularly when a gate voltage is applied. This is partially expected given the performance of the Cartesian mapping for the transition probability and energy spectrum of a simple 2-spin orbital and 4-spin orbital models.[41]

## B. Anderson impurity model

The time-dependent total current for the Anderson impurity model obtained from the Cartesian mapping is shown in Figure 2 for two temperatures $T = \Gamma$ (left panels) and $T = \Gamma/5$ (right panel). Each panel contains results for a set of different source-drain voltages. In all cases, the current saturates at high values of $V_{sd}$, as it should.

The behavior of the current is qualitatively different as we change the onsite repulsion $U$. For small values $U$ the current resembles that of the non-interacting case, shown in Figure 1. As $U$ increases, a pronounced oscillation in the current at early times is observed even at small source-drain voltage, signifying the repulsion between electrons of different spins. This is translated to a suppression of the steady-state current (inferred from the limit $\Gamma t/\hbar \to 5$) at small bias voltages. The suppression is more pronounced at lower temperatures.

Comparing the results at different temperatures for a given source-drain bias, we find that at low values of $V_{sd}$ the steady state current is higher for the higher temperatures for large values of $U$. Above a blockade voltage this behavior is reversed and the steady state value of the current at low temperatures is larger than the corresponding value at high temperatures. This crossover behavior captured by the quasi-classical approximation (and to a lesser extent by the action-angle mapping not shown here) is consistent with the quantum mechanical result. In fact, this is one of the hallmarks of Coulomb blockade, as further discussed below.

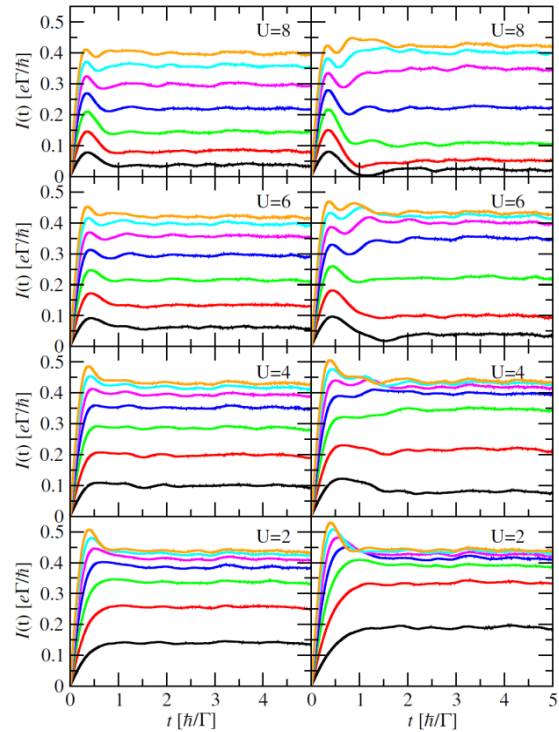

Figure 2: The total current as a function of time for the Anderson impurity model obtained from the Cartesian mapping. Left panels show the results for $T = \Gamma$ and right panels for $T = \Gamma/5$. Black, red, green, blue, magenta, cyan and orange correspond to $eV_{sd} = 2, 4, 6, 8, 10, 12$ and $14$ in units of $\Gamma$, respectively.

In Figure 3 we compare the steady state values of the current calculated from the quasi-classical mappings to quantum mechanical results obtained within an equation of motion approach to the nonequilibrium Green's function formalism, where all 2-body correlations between the system operators were included.[46] This approach is known to provide accurate results in the present regime of parameters and temperatures.[47] In fact, it can also capture the Kondo resonances when higher order correlations in the leads are included.[48]

When the onsite interaction is turned on the quasi-classical Cartesian mapping is not as accurate as it was in the non-interacting case (Figure 1). However, it captures qualitatively the dependence of the current on the source-drain bias for both temperatures studied. At the higher temperature, we find that the current increases roughly linearly with $V_{sd}$ before it bends and saturates at high values of $V_{sd}$. The quasi-classical Cartesian mapping performs better at low values of $V_{sd}$ and at higher values of $U$. It slightly underestimates the current at saturation. In this regime of parameters the action-angle mapping also performs well, particularly for small values of $U$.

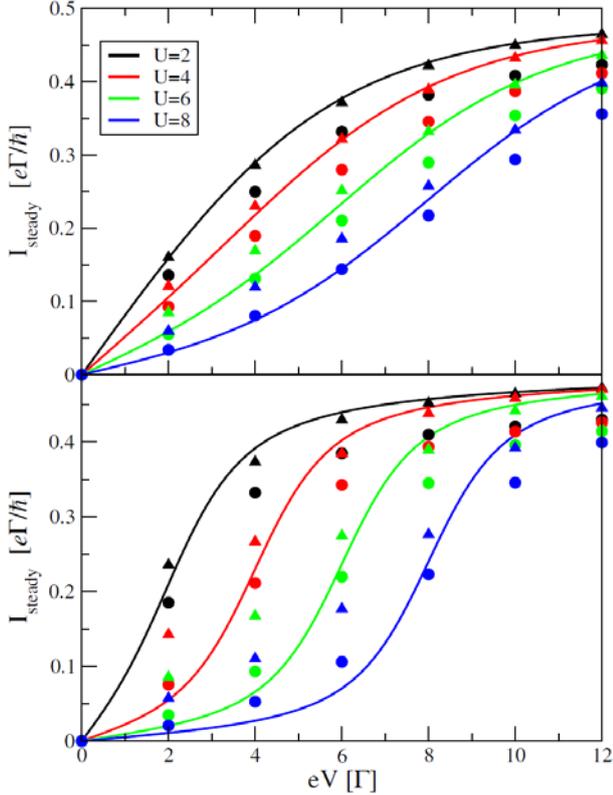

Figure 3: Steady state values of the current for the Anderson impurity model. Upper and lower panels are for $T = \Gamma$ and $T = \Gamma/5$, respectively. Solid lines represent quantum mechanical results based on a nonequilibrium Green's function approach, circles correspond to results of the quasi-classical Cartesian mapping, and triangles are the results of the action-angle mapping of Ref. 13, both calculated with $\lambda_1 = \lambda_2 \equiv \lambda = 20$.

When the temperature is reduced, the current assumes an "S"-shape voltage dependence, with a blockade that increases with $U$ The mid-point in the current versus bias voltage occurs when $V_{sd} = U$. Below this value, the current is blocked and only when the bias increases above $V_{sd} = U$, a conducting channel opens up and the current increases rapidly. This Coulomb blockade is qualitatively captured by the quasi-classical Cartesian mapping, and, to a lesser extent, by the action-angle mapping. Importantly, the Cartesian mapping captures the blockade and the overall shape of the current versus bias voltage. Again, the Cartesian approach slightly underestimates the current as it levels off.

It is important to note that the special treatment of the two-body terms as described in Section II.B is essential to describe the blockade phenomena. In fact, when we employ the mapped Hamiltonian given by Eq. (4) rather than that given by Eq. (6), the quasi-classical approach fails to reproduce the Coulomb blockade and the current increases even below $V_{sd} < U$. The effect of changing $\lambda_1$ and $\lambda_2$ on the steady state current is shown in Figure 4 for the special case $\lambda_1 = \lambda_2 \equiv \lambda$. As is clearly evident, only when $\lambda$ is sufficiently large, a clear blockade is developed. The results converge and change very slightly at $\lambda > 20$.

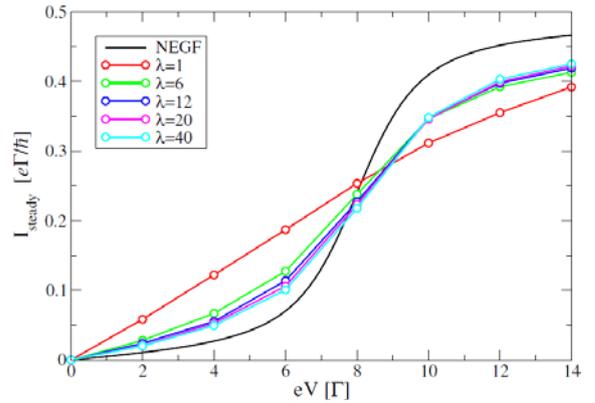

Figure 4: Steady state current versus the source-drain bias voltage for different values of $\lambda_1 = \lambda_2 \equiv \lambda$ calculated from the quasi-classical Cartesian mapping. The remaining parameters are: $U = 8\Gamma$ and $T = \Gamma/5$.

## IV. CONCLUSIONS

We have presented an approach based on a Cartesian mapping of the many-electron Hamiltonian[41] to calculate the nonequilibrium properties of the resonant level model and the Anderson impurity model. The mapping keeps track of the anti-commutation relation of the creation/annihilation operators, required for Fermi-Dirac statistic. It also provides a suitable framework for a quasi-classical approximation which accounts for the correct thermodynamics at $t = 0$ and for a semiclassical initial value treatment, which will be the subject of future study.

The approach provides excellent agreement for the resonant level model (non-interacting limit) in comparison to exact quantum mechanical calculations for a wide range of model parameters, including different temperatures and bias voltages. It also captures quantitatively the behavior of the current at different gate voltages, correcting for the flaws of our previous mapping based on the Miller-White[37] action-angle approach.

When we turn on the two-electron interactions the approach is no longer numerically exact. It performs quite well at high temperatures (but so does the action-angle mapping of Ref. 13). To capture the Coulomb blockade at lower temperatures and higher values of the onsite repulsion $U$, we have modified the two-body mapped interaction, using identities of creation/annihilation operators. The modified Hamiltonian captures the Coulomb blockade qualitatively and more importantly, the temperature dependence of the current below and above the blockade.

## Acknowledgments

We would like to thank Guy Cohen for fruitful discussions. This work was supported by the National Science Foundation Grant No. CHE-1148645 and by the Director, Office of Science, Office of Basic Energy Sciences, Chemical Sciences, Geosciences, and Biosciences Division, U.S. Department of Energy under Contract No. DE-AC02-05CH11231, by the FP7 Marie Curie IOF project HJSC, and by the US-Israel Binational Science Foundation. TJL is grateful to the Azrieli Foundation for the award of an Azrieli Fellowship. We also acknowledge a generous allocation of supercomputing time from the National Energy Research Scientific Computing Center (NERSC) and the use of the Lawrencium computational cluster resource provided by the IT Division at the Lawrence Berkeley National Laboratory.